\documentclass[aps,pre,twocolumn,groupedaddress,showpacs,
preprintnumbers,eqsecnum,
]{revtex4}

\usepackage{graphicx}
\usepackage{bm}

\begin{document}

\author{Joseph Rudnick}
\affiliation{Department of Physics and Astronomy, UCLA, Box
951547, Los Angeles, CA 90095-1547}
\author{David Jasnow}
\affiliation{Department of Physics and Astronomy, University of
Pittsburgh, Pittsburgh, PA 15260}

\date{\today}

\title{Cohesive energy, stability and structural transitions in
polyelectrolyte
bundles}

\begin{abstract}
A lattice of uniformly charged, infinitesimally thin, rods decorated
with an ordered array of counterions exhibits anomalous behavior as
the spacing between the rods is varied.  In particular, the counterion
lattice undergoes a sequence of structural shearing, or ``tilting,''
phase transformations as the spacing between the rods decreases.  The
potential implications of this behavior with respect to the packaging
of biologically relevant polyelectrolytic molecules are commented
upon.
\end{abstract}

\maketitle

\section{Introduction}

Polyelectrolyte chains naturally repel each other since they carry an
overall electric charge.  That these chains will nevertheless form
condensed phases in the presence of oppositely charged counterions has
been known for some time.  It has also become clear that this
condensation results from some form of organization of the
counterions, either dynamical~\cite{oosawabook,liu} or essentially
static, in the form of a counterion lattice~\cite{shk}.  In both
cases, the attraction between oppositely-charged regions overcomes
Coulomb repulsion between the bare chains and leads to a net
attraction.  Correlations in the counterion system are crucial, since
it is well-known that mean field theory, represented by solution of
the Poisson-Boltzmann system of equations, cannot yield an attraction
between like-charged 
rods~\cite{oosawabook,poisson-boltz,oosawa,ohnishi}.

The physics of polyelectrolytes is relevant to biological systems.
For example, double helix DNA will, under certain conditions,
organize into a condensed state in which it self-assembles into
bundles of densely packed parallel rods~\cite{b1,poisson-boltz}.
An important venue for this condensation is within the heads (or
\emph{capsids}) of various viruses~\cite{hen}.  This organization
occurs in spite of the fact that ``naked'' DNA carries a strong
negative charge---one excess electron per phsophate group on the
backbone~\cite{wat}.  DNA has also been observed to form condensed
liquid-crystal-like phases in the presence of polyvalent
counterions~\cite{livolant1,livolant2}.  In addition,
counterion-mediated formation of actin bundles has also been
observed~\cite{angelini}.  Interestingly, in this last case, there
is evidence of a kind of counterion lattice, in the form of a
charge-density-wave-like modulation along the axis of the
condensed actin filaments.  In such a lattice, the ``counterions''
in the lattice actually consist of clusters of individual
counterions, and can therefore not be represented as point
particles.  The question of an ionic lattice specifically in the
case of condensed DNA remains an open one, since there is, as yet,
no experimental indication of such an organization of counterions.
Furthermore, theoretical estimates indicate that a 3D structure
requires counterion valences, $Z$, greater than about
6~\cite{shk}.  A lattice of point-like counterions in condensed
biological rodlike molecules thus remains a conjecture, rather
than an established fact.  Nevertheless, it represents an
sufficiently interesting and potentially important construct
within biological and polyelectrolyte physics that its properties
merit investigation.

In this paper we analyze more carefully the cohesive energy of the
background of negatively charged rods along with a neutralizing,
crystalline counterion system.  We assume the counterions form a
\emph{modified} face-centered cubic (fcc) crystal.  Our principal
result is that, as the rod density increases, the counterion system
maintains stability by undergoing structural transitions to ``tilted''
lattices.  More precisely, we have located two symmetry-breaking
structural transitions: the first is of the three-state Potts type and
is therefore weakly first order.~\cite{amit} The second is a
continuous transition in the Ising universality class.  A fluctuation
analysis indicates that the transitions are mean field in nature. 
This reflects the dominating influence of long-range Coulomb
interactions between the unscreened charges that are bound to the
rods.  At the end we speculate on how the singularities associated
with such transitions could be relevant to the physics of packaging
and other bundling phenomena in biology.  It is worth noting that
structural transitions have been recently observed in DNA-dendrimer
complexes \cite{dendrimers}.  These transitions do not appear to
correspond to the specific ones we discuss.  However, the notion of
structural transitions in complexes of rods and localized charges
clearly has an experimental as well as a theoretical basis.

We also touch on, but do not discuss in detail, the counterion
``melting'' transition expected when the rods are far apart.  This
leads to a ``counterion liquid.''  The attraction between rods in such
a state has been extensively explored by Ha and Liu,
\cite{liu,liu1,liu2}.  The melting transition is continuous and can be
shown to be in the universality class of the three-dimensional $XY$
model; see, e.g., \cite{gruner,girault,monceau,brill}.  A brief review
of the arguments leading to this conclusion appears in an appendix.

An outline of this paper is as follows.  The following section
contains the characterization of the lattice and the Coulomb energy
calculations.  Section~\ref{sec:melting} addresses ``melting'' of the
counterion lattice, while Section~\ref{sec:concl} contains concluding
remarks.  A series of appendices address some technical issues. 
Appendix A contains a calculation of the Coulomb energy of a lattice
of infinite, uniformly charged rods.  Appendix~B provides the
derivation of the term in the energy quadratic in displacements of
counterions from their lattice sites (i.e., harmonic ``phonon"
dispersion relation).  Appendix~C provides some heuristics on the
``melting'' of the counterion lattice.

\section{Lattice of rods and counterion energy}
\label{sec:latt-and-coul}
The ``lattice'' consisting of a set of negatively charged rods and
attached polyvalent counterions of charge $+Ze$ will be treated as a
simple ionic lattice.  At the first stage of approximation, the fixed
charge of the rods will be replaced by a uniform negatively charged
background (as in the jellium model of interacting electrons
\cite{ziman}).  The counterions will be assumed to form a
three-dimensional lattice that generalizes the close-packing
arrangement in a face-centered cubic (fcc) crystal.  Given that the
rods are infinitesimally thin, there is no prospect of a
two-dimensional arrangement of charges on the surface of any of them
\cite{shk}.  As has been noted in the literature~\cite{A&M}, such a
lattice can be constructed starting from a two-dimensional hexagonal
close-packed structure, via the introduction of three sublattices. 
Our generalization makes use of this construction.  No significant
error is introduced by our neglect of the explicit contribution of the
rods themselves to electrostatic interactions.  Dimensional
considerations and detailed calculations (see Appendix
\ref{app:roden}) lead to the conclusion that the Coulomb energy due to
interactions between charged rods consists of two contributions, the
first independent of the separation between rods and the second going
as the logarithm of the separation between rods, assuming overall
charge neutrality.  Given that the counterions are forced to sit on
the rods, the principal outcome of the rod-counterion interaction
arises from the overall charge neutrality enforced by the charge on
the rods.

\subsection{Structure of the rod-counterion lattice}

\begin{figure}[htb]
\includegraphics[height=2in]{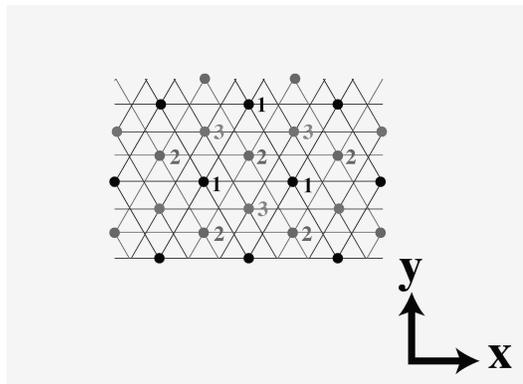}
\caption{Schematic of the counterion lattice, seen end-on with
respect to the rods that support it. The rods are hexagonally
close-packed, and the three sublattices occupied by the counterions
are indicated.}
\label{fig:latt1}
\end{figure}

\begin{figure}[htb]
\includegraphics[height=2in]{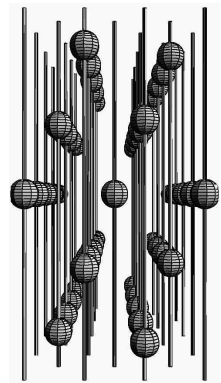}
\caption{A view of the lattice perpendicular to the axis of the
rods. The spheres represent the counterions. Scanning from left to
right, one observes the three sublattices.} \label{fig:lattice}
\end{figure}

Imagine a large bundle of hexagonally close-packed pencils or
rods. Viewed end-on, the rods lie on three triangular sublattices,
as depicted in Fig.~\ref{fig:latt1}. Now pass planes at regular
spacing $l_v$ perpendicular (initially) to the rods. Where the
first plane intersects the rods on the first sublattice, place
counterion charges. Likewise treat the second and third planes at
their intersections with rods on the second and third sublattices,
respectively. Repeat this pattern with subsequent planes, thereby
building an infinite three-dimensional modified fcc lattice.
Figure \ref{fig:lattice} displays the lattice from a viewpoint
perpendicular to the axis of the rods.  The three counterion
sublattices are evident if one scans the figure from left to right.
If $l_v$ is adjusted appropriately relative to the lattice constant of
one triangular sublattice, $l_h$, the counterion charges themselves
sit on a true fcc lattice.

The three primitive vectors of the Bravais lattice of counterions
so constructed are
\begin{eqnarray}
\vec{a}_{1} & = & 
\frac{1}{2}l_{h}\hat{x}+\frac{\sqrt{3}}{2}l_{h}\hat{y}
+l_{v}\left(g_{x}+g_{y}\sqrt{3}\right)\hat{z}  \label{a1} \\
\vec{a}_{2} & = &
-\frac{1}{2}l_{h}\hat{x}+\frac{\sqrt{3}}{2}l_{h}\hat{y}
+l_{v}\left(-g_{x}+g_{y}\sqrt{3}\right)\hat{z}  \label{a2} \\
\vec{a}_{3}& = & \frac{1}{\sqrt{3}}l_{h}\hat{y}+l_{v}\left( 1 
+\frac{2}
{\sqrt{3}} g_{y} \right)\hat{z}  \label{a3}
\end{eqnarray}

The dimensionless vector $\vec{g} = \hat{x}g_{x} + \hat{y}g_{y}$
encodes the possibility of tilting the planes of counterions.  The
corresponding primitive vectors of the reciprocal lattice,
$\vec{b}_{i}$, are constructed in the standard way \cite{A&M}, so that
the relationship between the two sets of primitive vectors is $
\vec{a}_{i} \cdot \vec{b}_{j}=2\pi \delta _{i,j} \, .$ To recover the
true fcc lattice, the \emph{aspect ratio}, $r \equiv l_{h}/l_{v}$, is
taken to be $\sqrt{3/2}$ and $\vec{g}$ is set equal to zero.  The
aspect ratio will play the role of a control parameter in what
follows.

Assume overall charge neutrality of the rod-counterion system, so that
there
is exact cancellation between the mean charge per unit volume of the
counterion lattice and the uniformly charged negative background
provided by
the rods. We make use of the Ewald method for the evaluation of the
Coulomb
sum~\cite{A&M,ewald,ziman}. The ``reduced'' Coulomb energy is defined 
as
\begin{eqnarray}  \label{Ecoul}
\lefteqn{E_{\mathrm{Coul}}/(Z^{2}e^{2}/2l_{v})} \nonumber \\ & \equiv 
&
\tilde{E}_{\mathrm{Coul}} \nonumber \\  &=&  \sum_{\vec{n}\neq
0}\frac{l_{v}}{|n_{1}\vec{a}_{1}+n_{2}\vec{a}
_{2}+n_{3}\vec{a}_{3}|}-S
\end{eqnarray}
where $\vec{n}$ is a triplet of integers $(n_{1},n_{2},n_{3})$, and 
the
subtraction $S$ represents the compensating interaction with the
smeared out
negatively charged background.

In our calculation, we assume that the distance between ``planes'' of
ions
is fixed at $l_{v}$. The effect of a compression of the lattice of
polyelectrolytic rods is to decrease the spacing, $l_{h}$.

\subsection{The Coulomb sum}
\label{sub:coul}
Using Ewald summation techniques \cite{ewald,A&M,ziman}, one
generates an expression for the Coulomb energy that can be expressed
in terms of the sum of four terms.  Those terms are
\begin{equation}
-\frac{2}{\sqrt{\pi}}\alpha^{1/2}\left(l_{h}^{2}l_{v}\right)^{-1/3}
    \label{cont1}
\end{equation}
\begin{eqnarray}
\lefteqn{\sum_{\vec{n} \neq
0}\frac{1}{|n_{1}\vec{a}_{1}+n_{2}\vec{a}_{2} + n_{3}\vec{a}_{3}|}}
\nonumber \\ && \times \left( 1 - \mathop{\rm erf}
\left(|n_{1}\vec{a}_{1}+n_{2}\vec{a}_{2} +
n_{3}\vec{a}_{3}|\alpha^{1/2} \left(l_{h}^{2} l_{v}\right)^{-1/3}
\right) \right) \nonumber \\
    \label{cont2}
\end{eqnarray}
\begin{equation}
-\frac{\pi}{v} \alpha^{-1} \left( l_{h}^{2}l_{v}\right)^{2/3}
    \label{cont3}
\end{equation}
\begin{equation}
\frac{4 \pi}{v} \sum_{\vec{m} \neq 0}\frac{e^{-|m_{1}\vec{b}_{1} +
m_{2}\vec{b}_{2} +m_{3}\vec{b}_{3}|^{2}\alpha^{-1}\left(l_{h}^{2} 
l_{v}
\right)^{2/3}/4}}{|m_{1}\vec{b}_{1} +
m_{2}\vec{b}_{2} +m_{3}\vec{b}_{3}|^{2}}
    \label{cont4}
\end{equation}
The quantity $\alpha$ in the expressions above is an
adjustable parameter, which is ideally set equal to a value that
maximizes convergence of the Ewald sums in (\ref{cont2}) and
(\ref{cont4}). A close-to-optimal choice is $\alpha= 4 \pi$.

\begin{figure}[htb]
\includegraphics[height=2.5in]{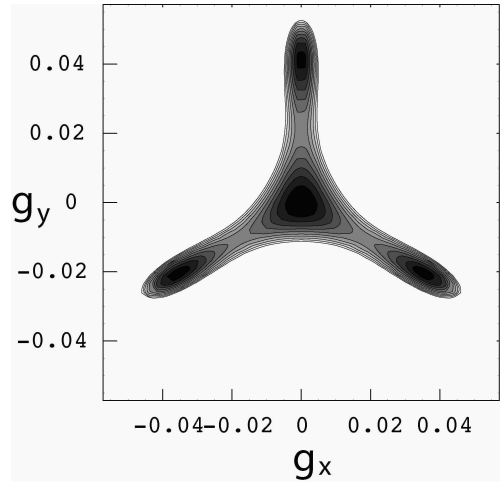}
\caption{Contour plot of the Coulomb energy of the generalized fcc
counterion lattice in the $g_{x}-g_{y}$ plane at
$r = r_{a} \approx 1.1$.
The four minima, including the one at $\vec{g}=0$,
are all of equal depth.}
\label{fig:firstorder}
\end{figure}

The mean-field phase diagram of the lattice can be determined by
examining the dependence of the Coulomb energy on the aspect ratio,
$r$, and the tilt vector, $\vec{g}$.  What one finds is that for $r$
greater than a threshold value $r_{a} \approx 1.1$ the Coulomb
energy is minimized when $\vec{g} =0$.  At this threshold value, three
minima lying symmetrically in the $\vec{g}$-plane represent equally
low energies.  This situation is illustrated in
Figure~\ref{fig:firstorder}.
When $r<r_{a}$, $\vec{g}=0$ no longer represents a global minimum of
the energy.  At $r=r_{s} \approx 1.097$, the local minimum at
$\vec{g}=0$ disappears.  In this sense, one can think of $r_{s}$ as a
``spinodal'' point.  Note the small difference between $r_{s}$ and
$r_{a}$.  The transition at $r_{a}$ is \emph{weakly} first order.

\begin{figure}[tbp]
\includegraphics[height=3in]{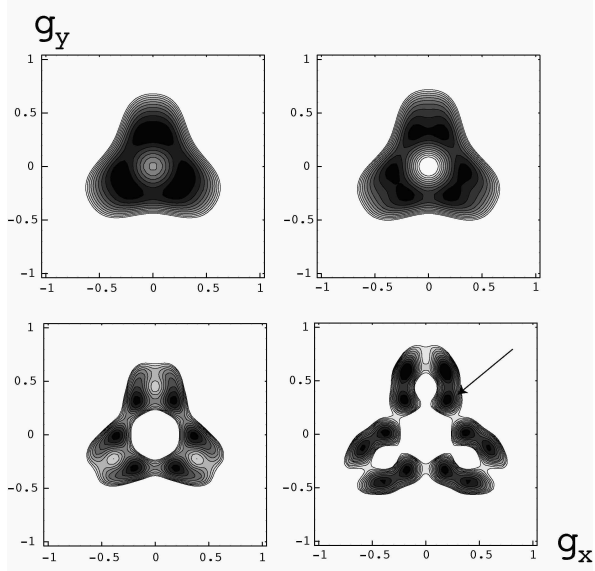}
\caption{The emergence of six minima in the Coulomb energy, and the
migration of those minima as the aspect ratio is decreased below the
threshold value $r_{b} = 0.801$.  Reading from left to right and top
to bottom, the values of $r$ are equal to 0.801, 0.693, 0.577 and
0.433, respectively.  The arrow in the bottom-right-hand figure points
to one of the six true minima of the energy.}
\label{fig:continuous}
\end{figure}

As $r$ is further reduced, corresponding to even closer packing of the
rods, a new structural transition is encountered, at which the three
minima each split into two new ones.  The aspect ratio, $r_{b}$, at
which this transition takes place is approximately equal to 0.801.
Contour plots illustrating the onset of this transition and the
evolution of the new minimum energy configurations are shown in
Figure~\ref{fig:continuous}.
That this transition is also a symmetry-breaking one is evident from
Fig.~ \ref{fig:continuous}.  Here, the transition is continuous, in
that the appearance of the six minima at lower aspect ratio is
simultaneous with the disappearance of the three minima associated
with values of $r$ greater than $r_{b}$.  At mean field level this
transition is of the standard second-order type.  Given the nature of
the symmetry-breaking, it is altogether reasonable to classify it as
an Ising-like, or $O(1)$, phase transition.  In the case at hand, the
aspect ratio plays the role of temperature.  This means that
singularities that one expects to find in the entropy and specific
heat of a thermal system will show up here in the form of
nonanalyticities in the dependence of the energy as a function of
spacing between rods, with direct consequences on packing forces.

In light of the continuous nature of the phase transition at $r=r_{b}$
and the weakness of the first-order phase transition at $r=r_{a}$, the
question of the effects of fluctuations is clearly relevant.  We
address this question with the use of coarse-grained effective
Hamilitonians.  Assume that the counterions are only allowed to move
along the rods.  Then, at the onset of the first-order transition, we
can express the fluctuations in the locations of the counterions in
terms of a scalar displacement field, $u(\vec{r})$ that quantifies
counterions displacments paralles to the rods.  The Coulomb energy can
then be written as an expansion in terms of $u$.  This energy is most
usefully expressed in terms of the Fourier transform of the
displacement field.  The most relevant terms in the expansion yield
the expression
\begin{eqnarray}
H & = & \sum_{\vec{k},Q}\left[|\vec{k}|^{2}\left(r-r_{s}\right) +
Ck^{4} +
AQ^{2} + B\frac{ Q^{2}}{k^{2}+Q^{2}} \right]  \nonumber \\
&&\times u(\vec{k},Q) u(-\vec{k},-Q)  \nonumber \\
&&+\sum_{\vec{q}_{1},\ldots,\vec{q}_{3}}w_{3}(\vec{q}_{1},\vec{q}_{2},\vec{q}%
_{3}) u(\vec{q}_{1})u(\vec{q}_{2}) u(\vec{q}_{3}) \delta_{\vec{q}_{1}
+\vec{q%
}_{2} + \vec{q}_{3}}  \nonumber \\
&& + \sum_{\vec{q}_{1},\ldots,\vec{q}_{4}}
w_{4}(\vec{q}_{1},\vec{q}_{2},%
\vec{q}_{3},\vec{q}_{4}) u(\vec{q}_{1})u(\vec{q}_{2}) u(\vec{q}_{3})
u(\vec{q%
}_{3})  \nonumber \\
&& \times \delta_{\vec{q}_{1} +\vec{q}_{2} + \vec{q}_{3}+ \vec{q}_{4}}
\label{quad1}
\end{eqnarray}
Here, we have split the three-dimensional wave vector $\vec{q}$ into a
two-dimensional vector, $\vec{k}$ in the $x$-$y$ plane and a
component, $Q$, in the $z$-direction.  The term with coefficient $B$
is the contribution to the quadratic energy reflecting the long-range
nature of the unscreened Coulomb interactions between counterions.  We
have inserted the ``stabilizing'' quartic term going as $k^{4}$, but
have ignored the inessential term proportional to $Q^{4}$, or the
cross term, proportional to $k^{2}Q^{2}$.  It is important to keep in
mind that the actual stabilization of the system results from the
shearing transition.

The higher order terms in the the effective Hamiltonian (\ref{quad1})
have
the following forms:
\begin{equation}
w_{3}(\vec{q}_{1},\vec{q}_{2},\vec{q}_{3}) = -C_{1}\left(k_{y}^{3} -
3k_{y}k_{x}^{2}\right)  \label{cubeterm1}
\end{equation}
and
\begin{eqnarray}
\lefteqn{w_{4}(\vec{q}_{1}, \cdots , \vec{q}_{4})}  \nonumber \\
&=& W(\vec{q}_{1}, \cdots , \vec{q}_{4})  \nonumber \\
&&+ B_{1} \left[ (\vec{k}_{1} \cdot \vec{k}_{2})( \vec{k}_{3} \cdot
\vec{k}%
_{4}) + \mbox{permutations} \right]  \nonumber \\
&&+ B_{2} Q_{1}Q_{2}Q_{3}Q_{4}  \nonumber \\
&& +B_{3} \left[ ( \vec{k}_{1} \cdot \vec{k}_{2}) Q_{3}Q_{4} + %
\mbox{permutations} \right] \, . \label{wform1}
\end{eqnarray}
Here,
\begin{eqnarray}
\lefteqn{W( \vec{q}_{1}, \cdots ,\vec{q}_{4})}  \nonumber \\
&=& v(\vec{q}_{1} + \vec{q}_{2}) + v(\vec{q}_{1} + \vec{q}_{3}) +
v(\vec{q}%
_{2} + \vec{q}_{3})  \nonumber \\
&& - v(\vec{q}_{1})-v(\vec{q}_{2})-v(\vec{q}_{3}) - v(-\vec{q}_{1} -
\vec{q}%
_{2} - \vec{q}_{3}) \, , \label{w0form}
\end{eqnarray}
where
\begin{equation}
v(\vec{q}) = \frac{Q^{4}}{|\vec{q}|^{2}} \, .  \label{vform}
\end{equation}
There are three distinct permuations of the indices in the first line
of the right-hand-side of Eq.  (\ref{wform1}) and six distinct
permutations in the last line of the right-hand-side of that equation.
The term $W(\vec{q}_{1},\ldots,\vec{q}_{4})$ appears to be the most
relevant contribution to the fourth-order coupling in the
Ginzburg-Landau-Wilson model appropriate to this system.  However, as
it turns out, the most important term in Eq.~(\ref {wform1}) is the
first one in the square brackets, going as $k^{4}$.

That the contributions to the energy associated with fluctuations
about the lattice have the forms shown  above can be established
through explicit evaluation of Coulomb-type lattice sums. Appendix
\ref{app:flucsums} outlines the calculation in the case of of the
quadratic terms and presents results for that and the fourth order
term that are obtained through explicit evaluation of those sums.

To perform an analysis of the effects of fluctuations on this weak
first-order transition, one can consider a ``Ginzburg criterion''
~\cite{ginzburg} applied to the one-loop contribution to the
``entropy'' of the system~\cite{ivanchenko}.  Recall that the leading
contribution to the mean field entropy goes as $r-r_{s}$ for $r
\rightarrow r_{s}+$.  Given the form of the of the quadratic term in
Eq.~(\ref{quad1}), and recalling that $s \sim \partial F/\partial r$,
with $F$ the free energy, we have the following expression for the
one-loop entropy,
\begin{eqnarray}
\lefteqn{s \sim \int \frac{k^{2}d^{3}q}{\left(r-r_{s}\right)k^{2}+
Ck^{4} +
AQ^{2} + B\frac{ Q^{2}}{k^{2}+Q^{2}}}}  \nonumber \\
&& \rightarrow \int \frac{k^{2}d^{2}kdQ}{\left(r-r_{s}\right)k^{2}
+ Ck^{4} + B\frac{ Q^{2}}{k^{2}}}.  \label{fluc1}
\end{eqnarray}
The expression on the right-hand-side of Eq.~(\ref{fluc1})
contains the important terms in the denominator. A variety of
methods exist for the evaluation of the integrals in this
expression. The essence of the results follows from a rescaling
$Q=k^{2}x$. Then, the integral to perform is
\begin{equation}
\int \frac{k^{2}d^{2}k dx}{\left(r-r_{s}\right) + Ck^{2}+Bx^{2}}
\label{fluc2}
\end{equation}
We now note that this has the same qualitative dependence on the
``reduced temperature,'' $r-r_{s}$, as the corresponding one-loop
integral of a five-dimensional $O(n)$ model with short range
interactions.  A further rescaling of the integration variables
produces a leading singularity proportional to $(r-r_{s})^{3/2}$
compared to the mean-field result $\propto r-r_{s}$.  This application
of the Ginzburg criterion shows that the transition is
\emph{unrenormalized}.  A calculation of the renormalization of the
fourth-order interaction leads to the same conclusion, namely that
fluctuations lead to a well-behaved change in the amplitude.  This
leads to the conclusion that, because of the long range dipole-dipole
interactions between the charges embedded on the rods, the first-order
shearing phase transition of the charge lattice at $r=r_a$ is
essentially mean field in nature.
Similar arguments, allowing for modifications in the form of
Eq.~(\ref{quad1}) by, for example, breaking rotational symmetry
in the $x-y$ plane, reveal that fluctuations don't modify the 
mean-field
\emph{continuous} shearing transition.  Technically, the transition
at $r=r_b$ is
Ising-like with effective dimensionality $d>4.$

\begin{figure}[htb]
\includegraphics[height=1.5in]{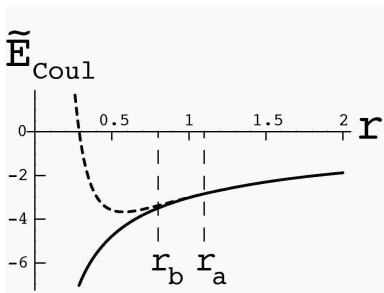}
\caption{
Solid line: the reduced Coulomb energy as a function of aspect
ratio in the vicinity of the first-order and continuous
transitions. Dashed line: the energy if the counterion
lattice is constrained not to tilt. The locations of the transitions
are indicated on the figure.}
\label{fig:engraph}
\end{figure}

The full results of the Coulomb energy calculations are
shown in Fig.~\ref{fig:engraph}. The analysis above reveals that
fluctuations don't qualitatively change the picture.

\begin{figure}[htb]
\includegraphics[height=1.5in]{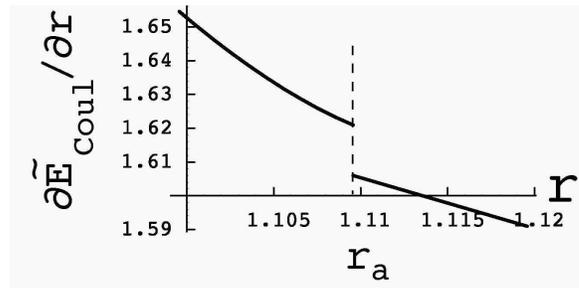}
\caption{A plot of $\partial \tilde{E}_{\mathrm{Coul}}
/\partial r$ in the immediate vicinity of $r=r_{a}$, illustrating the
discontinuity there, associated with the first order phase transition
to the sheared lattice.}
\label{fig:detail}
\end{figure}

Figure \ref{fig:detail} displays the (very small) discontinuity in the
derivative $\partial \tilde{E}_{\mathrm{Coul}} /\partial r$ at the
first order transition at $r=r_{a}$.

\section{``Melting'' of the counterion lattice}
\label{sec:melting}
The three-dimensional lattice of statically correlated counterions
being discussed can be reasonably expected to exist at low enough
temperatures and for sufficiently close packing of the lattice of 
rods.
At high temperatures and when the spacing between rods is
relatively large, the counterion lattice will not be stable against
thermal fluctuations. A variety of arguments lead to criteria for the
existence of this lattice. The simplest compares the (negative)
interaction energy of the lattice to $k_{B}T$. It is reasonable to
expect that the lattice will resist the disordering effects of
thermal fluctuations when
\begin{equation}
\left|\frac{Z^{2}e^{2}}{2l_{v}}\tilde{E}_{\mathrm{Coul}}\right| \geq 
k_{B}T
\label{eq:stab1}
\end{equation}
Assuming that the general condition, $Z^{2}e^{2}/(k_{B}T l_{v}) \gg 1$
for the existence of a ``Wigner crystal'' in the
couterion-polyelectrolyte system are met \cite{shk}, we expect the
crystal discussed here to be stable as long as
$|\tilde{E}_{\mathrm{Coul}}|$ is of order unity or greater. According 
to Fig.
\ref{fig:engraph}, this should be the case as for a range of aspect
ratios that encompass the two structural transitions on which this
paper has focussed.

An alternative estimation of the threshold at which melting of the
counterion lattice takes place is based on the Lindemann criterion
\cite{ziman}, according to which a lattice melts when
thermally-induced displacements are some fraction of the lattice
spacing.  This leads to approximate predictions for the melting
transition that we expect to be consistent with the
energy-{\em vs.}-entropy arguments leading to the criterion
(\ref{eq:stab1}) for the stability of the counterion lattice against
melting.

When melting takes place, it does so in the same way that charge
density waves disappear, that is to say, continuously, with
thermodynamic signatures that identify its universality class as that
of the $3DXY$ model.  Theoretical arguments and experimental
observations that justify this conclusion in the case of charge
density wave systems can be found in the literature
\cite{gruner,girault,monceau,brill}. Appendix \ref{app:melting}
contains a brief, heuristic argument for the nature of the melting
transition in the counterion system.

\section{Conclusions}
\label{sec:concl}
As noted above, the singularities in the dependence of the Coulomb
energy on aspect ratio have implications with respect to the
energetics of bundling or packaging polyelectrolytic molecules.  These
implications follow from the effective force exerted by the decorated
rods on each other, as determined by the ``pressure'', $p \equiv
-\partial E_{\mathrm{Coul}}/\partial r$.  The pressure (analogous to
the entropy in a thermal system as discussed) has a discontinuity at
$r=r_{a}$, and a discontinuous derivative at $r=r_{b}$.  The
singularities are not striking (recall Fig.  \ref{fig:detail}).  The
kinetics of bundling or packaging such molecules with crystalized
counterions are presumably affected by the Coulomb energy, and should,
in principle, reflect these singuarities.  What is more important,
however, is the possibility that a sequence of structural transitions
in the counterion lattice assists compression to high densities.  In
Fig.~\ref{fig:engraph} we have shown the Coulomb energy computed
directly from the Ewald sums in the vicinity of the first-order
transition at $r=r_{a}$ and the continuous transition at $r=r_{b}$. 
The dashed curve shows the energy if the lattice were constrained not
to tilt.  It is interesting to contemplate whether in some instances
Nature relieves the strong Coulomb repulsion via structural
transitions.

It is important to note that one cannot argue for structural
transitions as the {\em sine qua non} of polyelectrolyte packaging.
Specifically, in the packaging of DNA in tail-type bacteriophages, ATP
is known to provide the fuel for a packaging motor~\cite{hen,smith}.
Order of magnitude estimates suggest that $\sim 50-60 pN$ forces
generated by this motor~\cite{smith} suffice to overcome Coulomb 
forces
and compress a rod-like system to observed densities without
benefit of structural transitions.  Of course, in experimental
situations other repulsive energies besides Coulomb are also
involved.~\cite{riemer,odijk}

The present analysis does not suggest dramatic consequences in
actin bundling, or in the kinetics of packaging and/or infection
in a bacteriophage life-cycle. However, a continuous shearing
transition, if realized, should be accompanied by strong counterion
charge fluctuations, which could be susceptible to dynamic scattering
experiments.  The tilting, estimated to be a few percent at the
first-order transition, could potentially be detected by standard
diffraction techniques.

While, strictly speaking, we have shown that a shearing transition
ought to occur in a sufficiently closely-packed, constrained
polyelectrolytic system, we have not ruled out the possibility that
some other transition intercedes, preempting this particular
rearrangement as the system of rods is compressed.  One would have to
consider possibilities for ever larger unit cells in the rod lattice.
More importantly, we can only suggest the possibility that tilting or
other structural transition(s) occur as biologically relevant
polyelectrolytic molecules condense.
We hope this work will stimulate further experiments on, for example,
the
kinetics of DNA packaging, since such a transition could occur ``on 
the
fly''
as the ``spooling'' progresses.~\cite{odijk,spooling}
In other potential experiments, DNA or other bundles of varying 
density
could be prepared and probed statically.

We thank Prof.  A. B. Harris, Dr.  Ron Fisch and Prof.  Roger Hendrix
for helpful comments.  We are also grateful to Prof.  Robijn Bruinsma
for bringing Ref.  \cite{dendrimers} and the work on which it reports
to our attention.

\bibliographystyle{apsrev}
\bibliography{pack4}

\begin{appendix}

\section{The Coulomb sum for uniformly charged rods}
\label{app:roden}

As discussed in the text, the charges on the rods have been assumed to
be smeared out into a three-dimensional smooth background, analogous
to the jellium model used in discussions of electron-electron
interactions in a metal \cite{ziman}.  In this Appendix, we evaluate
the corrections to this approximation by now assuming that the
backbone charges are uniform {\em along the rods}.  The method
utilized to evaluate the alteration in the Coulomb energy that is
induced by this refinement in the model is a version of the Ewald
summation technique~\cite{ewald,A&M,ziman}.  The quantity we will
calculate is the potential energy of a charged test rod in the
presence of an array of uniformly charged rods, which are assumed to
be in a hexagonal close-packed arrangement.  At the end, the test rod
will be moved onto a rod of the lattice.  Before doing this, we
eliminate (``subtract") the interaction between the test rod charge
and the rod on which it eventually sits.

Because the rods are uniformly charged in the $z$-direction, the
potential at the two-dimensional location $\vec{r}$ due to a rod at
the origin is given by
\begin{equation}
\phi(\vec{r}) \propto \frac{1}{\pi} \int \frac{e^{i\vec{q}\cdot
\vec{r}}}{q^{2}}d^{2}q = \frac{1}{\pi} \int_{0}^{\infty}dt \left\{
\int e^{i\vec{q} \cdot \vec{r} -q^{2}t} d^{2}q \right\}
    \label{phi1}
\end{equation}
We split the $t$-integration into one from 0 to $T$ and another from
$T$ to $\infty$, where $T$ is, initially, arbitrary.  We suppose that
the distance between nearest-neighbor rods in the close-packed lattice
is $l_{h}$.  Given this, we will take $T \propto l_{h}^{2}$.  The
primitive vectors for this lattice can be taken to be
\begin{eqnarray}
\vec{a}_{1} & = &  l_{h}\hat{x}
    \label{a1r}  \\
\vec{a}_{2} & = &  \frac{l_{h}}{2}\hat{x} +
\frac{\sqrt{3}l_{h}}{2}\hat{y} ,
    \label{a2r}
\end{eqnarray}
and the volume of the primitive cell for this lattice is
\begin{equation}
v_{WS} = \frac{\sqrt{3}}{2}l_{h}^{2} .
    \label{vws}
\end{equation}
The corresponding reciprocal lattice vectors are
\begin{eqnarray}
b_{1} & = & \frac{2 \pi}{l_{h}}\left( \hat{x} - \frac{1}{\sqrt{3}}
\hat{y}\right)
    \label{b1}  \\
b_{2} & = &  \frac{4 \pi}{\sqrt{3}l_{h}} \hat{y} ,
    \label{b2}
\end{eqnarray}
while the volume of the primitive cell of the reciprical lattice is 
given by
\begin{equation}
v_{BZ} = \frac{8 \pi^{2}}{\sqrt{3}l_{h}^{2}}.
    \label{vbz}
\end{equation}
For the integration from $t=T$ to $t=\infty$, we make use of the
two-dimensional version of the Poisson sum formula:
\begin{equation}
\sum_{\vec{r} = \vec{R}_{i}} f(\vec{r}) = \frac{1}{v_{WS}} 
\sum_{\vec{Q}_{k}}
\int e^{i \vec{Q}_{k} \cdot \vec{r}}f(\vec{r}) d^{2}r ,
    \label{Psum}
\end{equation}
where $\{\vec{R}_{i} \}$ correspond to a hexagonal lattice, and where
the $\vec{Q}_{k}$ range over the reciprocal lattice.  We then have for
the contribution of this region of $t$-integration to the potential
energy of a point charge at the location $\vec{r}$
\begin{eqnarray}
\lefteqn{\frac{1}{\pi v_{WS}} \sum_{\vec{Q}_{k}} \int d^{2}r \int
d^{2}q e^{i\left( \vec{q}+ \vec{Q}_{k} \right) \cdot r }
\frac{e^{-Tq^{2}}}{q^{2}}} \nonumber \\ & = & \frac{4 \pi}{v_{WS}}
\sum_{\vec{Q}_{k}} \frac{e^{-TQ_{k}^{2}}e^{i\vec{Q}_{k} \cdot
\vec{r}}}{Q_{k}^{2}} .
    \label{part1}
\end{eqnarray}
The only tricky part here is the term for which $\vec{Q}_{k} =0$,
corresponding to the smeared-out portion of the distribution.  This
can be handled by a careful subtraction.
Because of the fact that $T \propto l_{h}^{2}$
and the fact that the $\vec{Q}_{k}$'s go as $l_{h}^{-2}$, terms in
(\ref{part1}) are independent of the spacing between rods, $l_{h}$.

The next portion of the integration is from 0 to $T$. Here, we sum
directly in real space. We find
\begin{eqnarray}
\lefteqn{\frac{1}{\pi} \sum_{\vec{R}_{l}} \int_{h}^{T}dt \int d^{2}q
e^{i\vec{q} \cdot \left( \vec{r}- \vec{R}_{l}\right)} e^{-tq^{2}}}
\nonumber \\
& = &  \sum_{\vec{R}_{l}} \int_{1}^{\infty} e^{-|\vec{r}-
\vec{R}_{l}|^{2}t/4T} \frac{1}{t} dt .
    \label{part2}
\end{eqnarray}
A careful subtraction eliminates the divergent contribution of term in
the sum in which $\vec{R}_{l} = \vec{r}$.  The divergence arises when
the limit $\vec{r} \rightarrow 0$ is taken, corresponding to placing
the test rod on one of the rods in the lattice.  One subtracts the
standard energy associated with the Coulomb interaction between the
test rod charge and the rod at the origin.  This subtraction leads to
the only term with any $l_{h}$ dependence, going as $\ln l_{h}$.  All
other dependence disappears because of cancellations between the
$l_{h}$-dependence of the $\vec{Q}_{k}$'s, the $\vec{R}_{l}$'s and
$T$.

The subtraction in (\ref{part1}) associated with the $\vec{Q}_{k} =0$
term is easily carried out, as is the subtraction in the case of the
term associated with a test rod position $\vec{r}$ directly on a rod,
in (\ref{part2}).  Choosing specifically
\begin{equation}
T = \frac{\alpha l_{h}^{2}}{4 \pi},
    \label{Tchoice}
\end{equation}
with $\alpha$ an arbitrary constant, one can perform the sums
indicated above, with the appropriate subtractions.  One finds the
following result for the Coulomb sum:
\begin{equation}
-2.78608 + 2 \ln l_{h}
    \label{sumresult}
\end{equation}
Note that the above result must be independent of the arbitrary 
constant
$\alpha$.  We are now in a position to include more precisely the
effect of the concentration of negative charge on the polyelectrolyte 
rods.
One must multiply the result (\ref{sumresult}) by
\begin{equation}
\sigma^{2} = \left(\frac{Ze}{l_{v}}\right)^{2}
    \label{sigma}
\end{equation}
where $\sigma$ is the linear charge density on the rods, $Z$ is the
valence of the condensed counterions, and $l_{v}$ is the spacing
between counterions on the rod. The above
multiplication will yield the energy of the rod lattice is per
unit rod length.

We are thus led to an expression for the contribution of the rod-rod
interaction to the Coulomb energy; it has a
non-trivial, but smooth dependence on the separation between rods.
This energy adds as a ``background'' term to the energies calculated
in the body of this paper.  A very similar calculation is used to
compute the interaction energy between the counterion lattice and
the uniformly charged rod lattice. Both contributions have no effect
on the structural transitions discussed in the body.

\section{Ewald sum for the second and higher order terms in counterion
displacements}
\label{app:flucsums}
The result for the energy cost of a distortion of the counterion
lattice follows straightforwardly from the expression for the
energy of a collection of interacting charges.  We start by writing
the two-particle interaction energy in terms of its spatial Fourier
transform,
\begin{equation}
V(\vec{r}) = \int d^{3}q v(\vec{q})e^{i \vec{q} \cdot \vec{r}}
\label{eq:enform}
\end{equation}
Then the positions of the counterions, $\vec{r}_{j}$, are expanded in
terms of displacements from the lattice sites, $\vec{u}(\vec{R}_j)$. 
Assuming that those displacements are entirely in the $z$-direction,
in line with our model in which the counterions are bound to the
charged polyelectrolye rods, the displacment of the $j^{\rm th}$
counterion from its equilibrium position on the counterion lattice
will be equal to $\hat{z} u(\vec{R}_j) = u_j \hat{z}$.  Expanding to
second order in the $u$'s, we obtain for the energy associated with
those distortions
\begin{equation}
\sum_{\vec{q}}\left[\sum_{\vec{Q}} \left( (q_{z}+ Q_{z})v(\vec{q}+ 
\vec{Q})
- Q_{z} v(\vec{Q}) \right)\right] u(\vec{q}) u(- \vec{q}) ,
\label{eq:secondorder1}
\end{equation}
where $u(\vec{q})$ is the spatial Fourier transform of the lattice
displacement $u_{j}$:
\begin{equation}
u(\vec{q}) = \sum_{j}u_{j}e^{-i\vec{q} \cdot \vec{R}_{j}}
\label{eq:uqdef}
\end{equation}

Given that the interactions are Coulomb, the sum of interest is
of the form
\begin{equation}
\sum_{\vec{Q}}\frac{\left(q_{z}+Q_{z}\right)^{2}}{|\vec{Q}+\vec{q}|^{2}} 
-
\sum_{\vec{Q}}\frac{\left(Q_{z}\right)^{2}}{|\vec{Q}|^{2}}
    \label{sum1pp}
\end{equation}
where the $\vec{Q}$'s are displacement vectors on the reciprocal
lattice. The primitive displacement vectors on this lattice are
$\vec{b}_{1}$, $\vec{b}_{2}$ and $\vec{b}_{3}$. The primitive
displacement vectors on the original lattice are $\vec{a}_{1}$,
$\vec{a}_{2}$ and $\vec{a}_{3}$. The relationship between the
$\vec{b}$'s and the $\vec{a}$'s is
\begin{equation}
\vec{b}_{1} = 2 \pi \frac{\vec{a}_{2} \times \vec{a}_{3}}{\vec{a}_{1}
\cdot \left(\vec{a}_{2} \times \vec{a}_{3}\right)}
    \label{b2p}
\end{equation}
and similarly for $\vec{b}_{2}$ and $\vec{b}_{3}$.  The primitive
volume in the reciprocal lattice is the volume of the first Brillouin
zone, given by
\begin{equation}
v_{BZ} = \left|\vec{b}_{1} \cdot \left(\vec{b}_{2} \times \vec{b}_{3}
\right) \right|
    \label{vbzp}
\end{equation}
The primitive volume in the real lattice is the volume of the
Wigner-Seitz cell, given by
\begin{equation}
v_{WS} = \left|\vec{a}_{1} \cdot \left(\vec{a}_{2} \times
\vec{a}_{3} \right) \right|
    \label{vwsp}
\end{equation}
Given the relationship between the $\vec{a}$'s and the $\vec{b}$'s,
the the following holds
\begin{equation}
v_{BZ} v_{WS} = \left(2 \pi \right)^{3}
    \label{vvpp}
\end{equation}
The Poisson sum formula in three dimensions takes the following form:
\begin{equation}
\sum_{\vec{Q}}f(\vec{Q}) = \frac{1}{v_{BZ}}\int d^{3}Q
\sum_{\vec{R}}e^{i \vec{R} \cdot \vec{Q}} f(\vec{Q})
    \label{Poisson1pp}
\end{equation}
The sum on the right hand side is over all lattice points on the real
space lattice.  One final relationship between the $\vec{a}$'s and the
$\vec{b}$'s is
\begin{equation}
\vec{a}_{i} \cdot \vec{b}_{j} = 2 \pi \delta_{i,j} .
    \label{adotbpp}
\end{equation}

In (\ref{Poisson1pp}) each $\vec{Q}$ is of the form
$m_{1}\vec{b}_{1}+m_{2}\vec{b}_{2} +m_{3}\vec{b}_{3}$, where the
$m_{i}$'s take integral values from $-\infty$ to $\infty$.  Similarly
the $\vec{R}$'s are of the form $n_{1}\vec{a}_{1} + n_{2}\vec{a}_{2} 
+ n_{3}
\vec{a}_{3}$, where the $n_{i}$'s range over all integers, as well.

Now, one can clearly write the expression in (\ref{sum1pp}) in the 
form
$f(\vec{q}) - f(0)$. Focus on $f(\vec{q})$ and write
\begin{equation}
\frac{\left(q_{z} + Q_{z}\right)^{2}}{\left|\vec{q}+\vec{Q}
\right|^{2}} = \int_{0}^{\infty}\left(q_{z} + Q_{z}\right)^{2} e^{-
| \vec{q}+\vec{Q} |^{2}t}dt
    \label{step1pp}
\end{equation}
As in Appendix A, the integral over $t$ splits into an integral from 0
to $T$ and from $T$ to $\infty$, where, in this case, we choose
\begin{eqnarray}
T &=& \beta \frac{\pi}{\left(v_{BZ}\right)^{2/3}} \nonumber \\
&=& \beta \frac{\left( v_{WS}\right)^{2/3}}{4 \pi} \,
\label{Tdefpp}
\end{eqnarray}
where $\beta$ is arbitrary.
For the integral from $T$ to $\infty$, one finds
\begin{eqnarray}
\lefteqn{\sum_{\vec{Q}}\frac{\left(q_{z}+ Q_{z}\right)^{2}}{\left| 
\vec{q} +
\vec{Q} \right|^{2}}e^{- \pi \beta |\vec{q}+
\vec{Q}|^{2}/v_{BZ}^{2/3}}} \nonumber \\ &=&
\frac{q_{z}^{2}}{\left|\vec{q}\right|^{2}}e^{ - \beta \pi
|\vec{q}|^{2}/v_{BZ}^{2/3}} \nonumber \\ && + \sum_{\vec{Q} \neq
0}\frac{\left(q_{z}+ Q_{z}\right)^{2}}{\left| \vec{q} + \vec{Q}
\right|^{2}}e^{- \pi \beta |\vec{q}+ \vec{Q}|^{2}/v_{BZ}^{2/3}}
    \label{sum2pp}
\end{eqnarray}

The Poisson sum formula is applied to the integration over $t$ from 0
to $T$.  This leads to the following expression
\begin{widetext}
\begin{eqnarray}
\int d^{3}Q
\left\{\frac{1}{v_{BZ}}\sum_{\vec{R}}\int_{0}^{\beta \pi
/v_{BZ}^{2/3}= \beta v_{WS}^{2/3} / 4
\pi}e^{i\vec{Q} \cdot
\vec{R}-|\vec{q}+\vec{Q}|^{2}t}\left(q_{z}+Q_{z}\right)^{2} dt
\right\}
    \label{sumbpp}
\end{eqnarray}
\end{widetext}
The integral in brackets is evaluated by introducing the generating
function
\begin{equation}
\frac{1}{v_{BZ}}\sum_{\vec{R}}\int_{0}^{ \beta v_{WS}^{2/3} / 4
\pi}e^{i\vec{Q} \cdot
\vec{R}-|\vec{q}+\vec{Q}|^{2}t + \kappa \hat{z} \cdot \left(\vec{q} +
\vec{Q} \right)} dt .
    \label{sumcpp}
\end{equation}
One obtains (\ref{sumbpp}) from (\ref{sumcpp}) by taking the second
derivative with respect to $\kappa$, and then setting $\kappa =0$.
For non-zero $\vec{R}$, the integral over $\vec{Q}$ is taken easily
enough.  Completing squares, one is left with
\begin{equation}
\frac{1}{v_{BZ}} \int_{0}^{\beta v_{WS}^{2/3}/4 \pi}
\left(\frac{\pi}{t}\right)^{3/2}e^{-i\vec{R} \cdot \vec{q}}
e^{\frac{1}{4t}\left[i\vec{R} + \kappa \hat{z}\right]^{2}}dt .
    \label{newintpp}
\end{equation}
Rescaling the integration variable and taking the requisite second
derivative with respect to $\kappa$ yields
\begin{equation}
\beta^{-3/2} \int_{1}^{\infty}t^{-1/2} e^{-i\vec{R} \cdot
\vec{q}}e^{- \pi R^{2}t/\beta v_{WS}^{2/3}}\left[ -\frac{\pi
R_{z}^{2}t^{2}}{\beta v_{WS}^{2/3} } + \frac{t}{2} \right] dt .
    \label{newerintpp}
\end{equation}
There is a singular contribution to the total sum from the term in
which $\vec{R} =0$. However, that term is independent of $\vec{q}$,
and is, therefore, cancelled when the total expression with $\vec{q}
=0$ is subtracted. The nonvanishing contribution is the sum over
non-zero $\vec{R}$'s of the expression in (\ref{newerintpp}). To this 
is
added (\ref{sum2pp}), and the result is $f(\vec{q})$ defined
through (\ref{sum1pp}).

\begin{figure}[htb]
\includegraphics[height=2in]{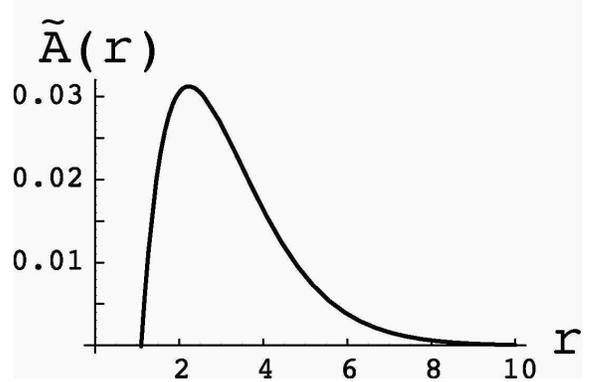}
\caption{The quantity $\tilde{A}(r) = 2A(r)/(Z^{2}e^{2}/k_{B}Tl_{v})$,
where $A(r)$ is the coefficient of $k^{2}$ in the quadratic energy of
distortion of the modified fcc lattice, quadratic in the displacement
field, as a function of the aspect ratio $r=l_{h}/l_{v}$.  In this
plot, the $z$-component of the wave vector $\vec{q}$ of the distortion
has been set equal to zero.}
\label{fig:quad1}
\end{figure}

Note that in all the above, the parameter $\beta$ has not been fixed.
It is, in fact, left undetermined, and may be set to speed convergence
of the sums.  Alternatively, it may be left as an internal check on
the procedure, since the final result must be independent of $\beta$.

To see what the sum developed above yields, we split the wave vecto
$\vec{q}$ as follows: $\vec{q} = \hat{z}Q + \vec{k}$, where the
projection, $\vec{k}$, lies in the $x$-$y$ plane.  Numerical results
are consistent with an energy quadratic in displacements that is
proportional to $k^{2}$ when $Q=0$ and $\vec{k}$ is small.  Figure
\ref{fig:quad1} shows the coefficient of
$k^{2}$ %
in that line as a function of the aspect ratio $r=l_{h}/l_{v}$.  Note
that the coefficient goes through zero at a value of $r$ that is close
to 1.  This corresponds to the ``spinodal instability'' lying below
the first-order transition at $r \approx 1.1$.  Near the spinodal the
harmonic spectrum takes the form shown
in Eq. \ref{quad1}.%
The coefficient $A(r)$ goes through a maximum as $r$ increases above
this value, and then decreases, tending to zero asymptotically as the
aspect ratio becomes large.  This latter tendency reflects the
weakening of the interaction that stabilizes the counterion lattice as
the distance between the rods grows in comparison to the distance
between neighboring charges on a rod.  In fact, it is not hard to
demonstrate that $A(r)$ will decay exponentially as $r$, the aspect
ratio,gets large.

To deal with terms that are third and fourth order in the
displacement field, the procedure is the same as the one
discussed above.  One is left with lattice sums to perform, and the
Ewald method leads to rapidly converging numerical algorithms.  The
results (\ref{cubeterm1})--(\ref{vform}) are obtained.  We find for 
the
coefficients $B_{1}$--$B_{3}$ in (\ref{wform1}), $B_{k} =
(Z^{2}e^{2}/2k_{B}Tl_{v}) \tilde{B}_{k}$, where
\begin{eqnarray}
\tilde{B}_{1} & = & 0.143
    \label{A1}  \\
\tilde{B}_{2} & = & 3 \times (-0.163)
    \label{A2}  \\
\tilde{B}_{3} & = & 0.010
    \label{A3}
\end{eqnarray}

\section{Counterion melting  is 3DXY}
\label{app:melting}

As the melting transition is approached from the ``counterion liquid''
side, we assume that there is an instability leading to a modulated
counterion charge density on each rod; then we allow for phase
fluctuations.  For the counterion density on each rod, we write
\begin{equation}
\rho(z) = A \cos \left({\cal Q} z + \phi(z) \right)
    \label{rhoform}
\end{equation}
Phase fluctuations disorder the charge density wave 
above and render the mean charge density on a rod
statistically uniform in the ``liquid'' phase.  The interaction
between counterions on a single rod will be of the form
\begin{equation}
\frac{1}{2} \int \int \rho(z) \rho(x^{\prime})V(z-z^{\prime}) dz \
dz^{\prime}
    \label{int1}
\end{equation}
Making use of (\ref{rhoform}), one obtains terms of the form
\begin{equation}
\int dz \int dz^{\prime} V(z-z^{\prime}) \exp \left[
i{\cal Q}(z-z^{\prime}) + i \left( \phi(z) - \phi(z^{\prime}) \right) \right]
    \label{int2}
\end{equation}
We now go to ``center of mass'' and ``relative'' co-ordinates. Let
\begin{eqnarray}
Z & = & \frac{z+z^{\prime}}{2}
    \label{Sig}  \\
{\cal Z} & = & \frac{z-z^{\prime}}{2}
    \label{Del}
\end{eqnarray}
The double integral in (\ref{int2}) is, then, proportional to
\begin{eqnarray}
\lefteqn{\int d Z \int d {\cal Z} V({\cal Z}) \exp \left[i {\cal Q} {\cal Z} 
+
2 i \phi^{\prime}\left( Z\right) {\cal Z} + \cdots \right]}
\nonumber \\ &= & \int d Z \int d {\cal Z} V({\cal Z}) \exp \left[i {\cal Q}
{\cal Z} \right] \nonumber \\ && \times \left( 1 + 2 i
\phi^{\prime}\left( Z\right) {\cal Z} + \frac{1}{2} \left( 2 i
\phi^{\prime}\left( Z\right) {\cal Z} \right)^{2} + \cdots \right)
\nonumber \\
& = & \int d Z \left[ v({\cal Q}) - 2i \phi^{\prime}(Z)
v^{\prime}({\cal Q}) + 2 \left( \phi^{\prime}( Z) \right)^{2}v^{\prime
\prime}({\cal Q}) \right] \nonumber \\
    \label{expand}
\end{eqnarray}
Now, let's assume that
\begin{equation}
v(q) \propto \int dz \exp[iqz] V(z) dz
    \label{vQ}
\end{equation}
has a minimum at $q={\cal Q}$. Then, $v^{\prime}({\cal Q}) =0$, and $v^{\prime
\prime}({\cal Q}) >0$. There is thus a contribution to the total energy going
as $\int \left( d \phi(z)/dz \right)^{2}dz$.

Now consider the interaction between rods. One expects that there 
will be
terms of the form
\begin{eqnarray}
\lefteqn{\int dz_{1} \int dz_{2} W(z_{1}-z_{2})} \nonumber \\ &&
\times \exp \left[ i {\cal Q} \left( z_{1}- z_{2}\right) + i \left(
\phi_{1}(z_{1}) - \phi_{2}(z_{2}\right) \right]
    \label{interaction}
\end{eqnarray}
The $\phi$'s are subscripted to make it clear that they refer to
different rods.  If one assumes that the interaction $W$ is
sufficiently short-ranged, which seems to be the case even for
unscreened Coulomb interactions because of the sinusoidal nature of
the assumed state, one reveals the essence by replacing
$\phi_{2}(z_{2})$ by $\phi_{2}(z_{1})$, and the integration in
(\ref{interaction}) becomes
\begin{eqnarray}
\lefteqn{\int dz_{1} \exp \left[ i( \phi_{1}(z_{1}) - \phi_{2}(z_{1}))
\right]} \nonumber \\ & & \times \int dz_{2} W(z_{1}-z_{2}) \exp
\left[ i {\cal Q}(z_{1}-z_{2}) \right] \nonumber \\ & & = w({\cal Q})
\int dz_{1} \exp \left[ i( \phi_{1}(z_{1}) - \phi_{2}(z_{1})) \right]
\nonumber \\
& & \rightarrow \int w({\cal Q}) \cos \left[ \phi_{1}(z_{1}) -
\phi_{2}(z_{1}) \right] dz_{1}
    \label{int3}
\end{eqnarray}

The result of this heuristic derivation is that there are terms in the
energy going as $(d\phi(z)/dz)^{2}$ and that there are also terms
going as $\cos (\phi_{1}(z)-\phi_{2}(z))$, where the subscripts refer
to near-neighbor rods.  The universality class for the transition in
this model is that of a $3DXY$ model. That the model is spatially
anisotropic does not influence the universality class.

\end{appendix}

\end{document}